\documentclass{article}

\usepackage[preprint]{spconf}

\usepackage[T1]{fontenc}
\usepackage[utf8]{inputenc}
\usepackage{amsmath,amssymb}
\usepackage{graphicx}
\usepackage{cite}
\usepackage{microtype}

\usepackage{booktabs}
\usepackage{multirow}
\usepackage{xcolor,colortbl}
\usepackage{comment}
\usepackage{pifont}

\usepackage{algorithm}
\usepackage{algpseudocode}
\usepackage{etoolbox}
\usepackage[hidelinks]{hyperref}
\AtBeginEnvironment{algorithmic}{\scriptsize}

\newcommand{\cmark}{\ding{51}}
\newcommand{\xmark}{\ding{55}}

\title{SwinDS-BWE: A Parameter-Efficient Swin-1D Lattice Generator with Decision-Science Discriminators for Speech Bandwidth Extension}

\hypersetup{
  pdftitle={SwinDS-BWE: A Parameter-Efficient Swin-1D Lattice Generator with Decision-Science Discriminators for Speech Bandwidth Extension},
  pdfauthor={Tarikul Islam Tamiti, Sajid Fardin Dipto, Luke Baja-Ricketts, David Vergano, Anomadarshi Barua},
  pdfsubject={Submitted to the 2027 IEEE International Conference on Acoustics, Speech and Signal Processing (ICASSP 2027)}
}

\name{Tarikul Islam Tamiti, Sajid Fardin Dipto, Luke Baja-Ricketts, David Vergano, Anomadarshi Barua}
\address{Department of Cyber Security Engineering, George Mason University, Virginia, USA}
\toappear{Submitted to the 2027 IEEE International Conference on Acoustics, Speech and Signal Processing (ICASSP 2027).}

\begin{document}
\ninept
\maketitle
\begin{abstract}

We propose SwinDS-BWE, a decision-science-inspired bandwidth extension (BWE) model with two coupled contributions: \textbf{(1)} Swin Transformers with lattice-style cross-stream interaction yield locality-aware modeling with linear-in-sequence per-window attention cost and reduce the prior AP-BWE generator size 0.5x from 33M to 17M. \textbf{(2)} Three novel lightweight decision-science-inspired discriminators augment AP-BWE’s performance: a \textbf{CVaR discriminator} tail-pools activations to emphasize worst high-frequency (HF) segments, a \textbf{Chance-Constraint HF discriminator} penalizes excessive HF energy via a differentiable barrier, and a \textbf{Multi-Criteria Utility discriminator} learns convex style weights over spectral criteria. SwinDS-BWE surpasses prior AP-BWE with a 30x smaller discriminator (42.3M vs. 1.36M) and higher fidelity on English and French datasets. 
This work shows that decision-science-inspired critics can supervise BWE to reduce discriminator size.

\end{abstract}


\vspace{-0.6em}
\section{Introduction}
\vspace{-0.4em}

Speech bandwidth extension (BWE) reconstructs high-frequency (HF) content that is missing after bandwidth compression, improving perceived clarity and benefiting downstream tasks, such as automatic speech recognition (ASR) and neural vocoders \cite{lu2024multi, su2021bandwidth}. 

Two practical bottlenecks surface when pushing BWE toward high-fidelity and deployment-ready operation. \textit{First}, \textbf{Generator Efficiency:} Many recent GAN models \cite{tagliasacchi2020seanet,mandel2023aero,hauret2023eben,liu2022neural,tamiti2025high} rely on parameter-heavy backbones or long-range/global mixing, whose cost scales poorly with sequence length, even though the hardest HF events are temporally local. \textit{Second}, \textbf{Supervision Mismatch:} Adversarial critics and reconstruction losses typically aggregate errors in a \textit{near-mean} sense over time \cite{kim2023hifi++, lu2024multi, sui2024tramba, 11464384}. This makes gradients dominated by easy vowel regions and can \emph{average away} rare HF failure events. Conversely, critics that reward generic ``sharpness'' can be gamed by excessive HF amplification, producing bright but unrealistic artifacts \cite{tamiti-etal-2026-cis}.

We address both issues by aligning (i) the generator’s inductive bias with the locality of HF structure, and (ii) adversarial supervision with the \emph{decision-relevant tail events} that improve perceived failures. Concretely, we propose \textbf{SwinDS-BWE}, an adversarial BWE system that couples a \textbf{Swin}-1D \cite{liu2021swin} lattice \cite{luo2020latticenet} generator with \textbf{D}ecision-\textbf{S}cience-inspired discriminators. The Swin-1D lattice generator replaces prior AP-BWE's \cite{lu2024towards} heavy amplitude--phase (AP) ConvNeXt ~\cite{liu2022convnet} refiners with \emph{shifted-window self-attention module along the time axis} (see Section \ref{subsec:gen}). This window attention limits computation to local neighborhoods, while alternating window-shifting enables cross-window information flow without global quadratic attention, reducing generator size without sacrificing performance. 

Moreover, we introduce three lightweight discriminators inspired by decision science to fix mean-aggregation failures: (i) A \textbf{\textbf{C}onditional \textbf{V}alue \textbf{a}t \textbf{R}isk \textbf{D}iscriminator (\textbf{CVaRD})} replaces mean pooling with tail pooling (TopK/CVaR-style) \cite{uryasev2000conditional} so that discriminator scores (and thus gradients) emphasize the \emph{worst} reconstructed segments rather than the \textit{average.} (ii) A \textbf{\textbf{C}hance-\textbf{C}onstraint HF \textbf{D}iscriminator (\textbf{CCD})} discourages pathological HF ``brightness'' by penalizing exceedance of a physically meaningful HF-energy ratio through a smooth barrier, acting as a differentiable feasibility regulator \cite{charnes1959chance}. (iii) A \textbf{\textbf{M}ulti-\textbf{C}riteria \textbf{U}tility \textbf{D}iscriminator (\textbf{MCUD})} computes interpretable spectral criteria (flatness/centroid/rolloff) and learns a convex utility (simplex weights) \cite{keeney1993decisions} that best separates real from fake, yielding a coherent signal rather than an unstable ad-hoc mixture. The critics are trained in lockstep \cite{durugkar2017generative} to prevent any single objective from dominating the adversarial game.

Here, for the first time, we frame BWE supervision into a decision problem: \textit{optimize not only the average frame, but the rare frames that matter most perceptually}. Empirically, this yields favorable performance, reducing overall model size substantially by 74\% compared to prior baseline AP-BWE (e.g., $72$M vs. $18.45$M) while maintaining strong perceptual scores across extension ranges. 
Moreover, our ablation study also supports the following contribution: \textit{decision-science critics provide complementary HF ``polishing'' that is most effective when paired with stabilizing global amplitude-phase feedback, while the Swin-1D generator improves perceived quality under the same discriminator feedback with markedly fewer parameters.} Source code will be released upon acceptance of this paper.


\begin{figure*}[htbp]
  \centering
\includegraphics[width=0.98\textwidth,height=0.165\textheight]{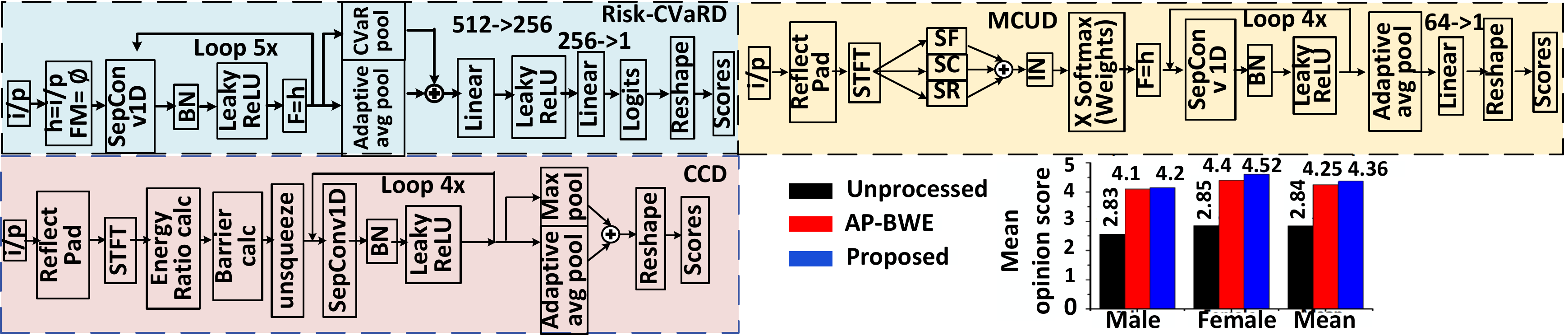}
\vspace{-01.0em}
  \caption{Our proposed CVaRD, CCD, and MCUD discriminators. (Right corner) Results of subjective MOS.}
  \label{fig:overall_architecture}
  \vspace{-01.8860em}
\end{figure*}

\vspace{-0.8em}
\section{SwinDS-BWE Architecture}
\label{sec:overview}
\vspace{-0.0em}


\vspace{-0.41em}
\subsection{Generator: Swin-1D with Lattice Coupling (see Alg.~\ref{alg:gen})}
\label{subsec:gen}

\vspace{-0.41em}


\,\,\,\,\,\,\,\,\, \textbf{a) Amplitude Phase (AP) lattice coupling:} The generator takes narrowband magnitude $M^{\mathrm{nb}}$ and phase $\Phi^{\mathrm{nb}}$ and parallely embeds them into magnitude and phase streams ($x$ and $y$) (\textbf{Rows \textcircled{\scriptsize 1}--\textcircled{\scriptsize 2}}). It then applies lattice-style \cite{luo2020latticenet} Swin blocks \cite{liu2021swin} twice for iterative refinement, as well as cross-stream information exchange via AP coupling (\textbf{Row \textcircled{\scriptsize 3}--\textcircled{\scriptsize 8}}). Within each block, phase is refined first via a Swin-1D Shifted Window Multi-Headed Self Attention (SW-MSA) block applied to $y$  (\textbf{Row \textcircled{\scriptsize 4}}), coupled into magnitude via learnable coefficients $(a_1,a_2)$ (\textbf{Row \textcircled{\scriptsize 5}}), the coupled magnitude is refined (\textbf{Row \textcircled{\scriptsize 6}}) via a Swin-1D, and the result is conditioned back into phase via learnable coefficients $(b_1,b_2)$ (\textbf{Row \textcircled{\scriptsize 7}}).  


\textbf{b) SW-MSA as an efficient temporal refiner:} Each refinement operator is a Swin-1D SW-MSA stack, which is invoked twice per lattice block: on the phase stream $y$ (\textbf{Row \textcircled{\scriptsize 4}}) and on the coupled magnitude state (\textbf{Row \textcircled{\scriptsize 6}}). Attention is computed within windows of size $W$ while alternating shifting of $\{0,W/2\}$ is used to enable cross-window propagation over successive blocks, yielding windowed complexity $O(TW)$ rather than $O(T^2)$ for global attention.


\textbf{c) Output heads: residual magnitude and complex phase prediction:} The wideband magnitude is predicted as a residual correction on the narrowband magnitude (\textbf{Row \textcircled{\scriptsize 9}}). Due to the noisiness of phase \cite{sfeclis2023investigating}\cite{yin2020phasen}, we do not use any residual connection in phase prediction. Wideband phase is produced by a complex-valued head via $\operatorname{atan2}(\cdot,\cdot)$ (\textbf{Row \textcircled{\scriptsize 10}}).

\vspace{-0.61em}
\begin{algorithm}[ht!]
\caption{Generator: Swin-1D with Lattice Coupling }
\label{alg:gen}
\begin{algorithmic}[1]\scriptsize
\Require $M^{\mathrm{nb}}, \Phi^{\mathrm{nb}}$; Swin window $W$ (shift $W/2$); two lattice blocks; learnable $(a_1, a_2, b_1, b_2)$ per block
\Ensure $M^{\mathrm{wb}}, \Phi^{\mathrm{wb}}$ (Wideband Magnitude and Phase)

\State $x \gets \mathrm{LN}(\mathrm{Conv1D}(M^{\mathrm{nb}}))$ \Comment{pre-project mag stream (conv + layer norm)}
\State $y \gets {\mathrm{LN}}(\mathrm{Conv1D}(\Phi^{\mathrm{nb}}))$ \Comment{{pre-project} phase stream (conv + layer norm)}
\For{$j=1$ to $2$} \Comment{two lattice blocks}
  \State $u \gets \mathrm{Swin1D}(y)$ \Comment{refine phase (2 blocks, shifted windows)}
  \State $m_1 \gets x + a_1 u;\ \ n_1 \gets a_2 x + u$ \Comment{AP coupling into magnitude}
  \State $v \gets \mathrm{Swin1D}(m_1)$ \Comment{refine coupled magnitude}
  \State $x \gets b_1 n_1 + v;\ \ y \gets n_1 + b_2 v$ \Comment{couple back into phase}
\EndFor
\State $M^{\mathrm{wb}} \gets M^{\mathrm{nb}} + \mathrm{Linear}(\mathrm{LN}(x))$ \Comment{pred mag + residual mag}
\State $\Phi^{\mathrm{wb}} \gets \operatorname{atan2}\left(\mathrm{Lin}_i(\mathrm{LN}(y)), \mathrm{Lin}_r(\mathrm{LN}(y))\right)$ \Comment{pred phase}
\State \Return $M^{\mathrm{wb}}, \Phi^{\mathrm{wb}}$ \Comment{form complex WB}
\end{algorithmic}
\end{algorithm}
\vspace{-0.82em}

\vspace{-01.3em}
\subsection{Decision-Science Discriminators and Terminology}
\vspace{-0.4em}
\label{subsec:disc}

Our decision-science discriminators are designed to be efficient and lightweight using depthwise-separable convolution (SepConv) \cite{howard2017mobilenets} as the core block. 
All decision-science critics are trained by hinge losses \cite{miyato2018spectral} to provide adversarial feature-matching feedback to the generator. Here, each critic outputs real/fake hinge logits $r,f\in\mathbb{R}^{B\times 1}$ (returned as $[r],[f]$) and intermediate feature maps $\{F_r\},\{F_f\}$ used for real/fake feature matching, respectively. 

\vspace{-0.7em}
\begin{algorithm}[ht!]
\scriptsize
\caption{CVaR Discriminator (CVaRD)} 
\label{alg:cvar}
\begin{algorithmic}[1]\scriptsize
\Require $y_{\mathrm{real}}, y_{\mathrm{fake}} \in \mathbb{R}^{B \times 1 \times T}$; tail fraction $\alpha$
\Ensure $[r], [f]$ and feature matching Feature Maps (FM) $\{F_r\}, \{F_f\}$

\State $(h_r, \{F_r\}) \gets \mathrm{SepConvBackbone}(y_{\mathrm{real}})$ \Comment{stride-2 SepConv stack}
\State $(h_f, \{F_f\}) \gets \mathrm{SepConvBackbone}(y_{\mathrm{fake}})$ \Comment{same forward on fake}
\State $k \gets \max(1, \lfloor \alpha \cdot T' \rfloor), \quad T' = \mathrm{len}_t(h_r)$ \Comment{tail size}
\State $z_r \gets \mathrm{CVaRPool}(h_r, k) + \mathrm{GAP}(h_r)$ \Comment{CVaRPool: avg of Top-$k$ over time (channel-mean)}
\State $z_f \gets \mathrm{CVaRPool}(h_f, k) + \mathrm{GAP}(h_f)$ \Comment{same summary on fake}
\State $r \gets \mathrm{MLP}(z_r)$ and $f \gets \mathrm{MLP}(z_f)$ \Comment{score head}
\State \Return $([r], [f], \{F_r\}, \{F_f\})$ \Comment{FM maps are backbone features}
\end{algorithmic}
\end{algorithm}

\vspace{-01.7em}
\subsubsection{\textbf{CVaR Discriminator (CVaRD): For tail-risk pooling for rare HF artifacts (see Algorithm~\ref{alg:cvar})}}
\label{subsec:disc_cvar}
\vspace{-0.41em}

Mean or average pooling \cite{lu2024towards} can cause adversarial feedback to fade away when HF artifacts are short and intermittent. The CVaRD complements \emph{mean-style} temporal aggregation with a \emph{tail-risk} (CVaR/Top-$k$) summary \cite{uryasev2000conditional} so that gradients can put the adversarial pressure in the worst reconstructed segments, along with retaining global context.  Given real/fake wideband waveforms 
$y_{\mathrm{real}},y_{\mathrm{fake}}$, a lightweight one-dimensional (1D) SepConv backbone produces a final temporal feature map $h$  for real/fake as $\{h_r,h_f\}$ and intermediate feature maps for feature matching, such as $(h_r,\{F_r\})$ and $(h_f,\{F_f\})$ \textbf{(Rows \textcircled{\scriptsize 1}--\textcircled{\scriptsize 2}}). Here, $h\in\mathbb{R}^{B\times C\times T'}$ where $B$ = batch size, $C$ = channel dimension, and $T'=\mathrm{len}_t(h_r)$ is the downsampled time length. For tail fraction $\alpha\in[0,1]$, we set the tail size $k=\max(1,\lfloor \alpha T' \rfloor)$ \textbf{(Row \textcircled{\scriptsize 3}}). To form a per-time evidence vector, we reduce channels, and define an empirical CVaR/Top-$k$ pool as $\mathrm{CVaRPool}(h,k)=\frac{1}{k}\sum_{t\in\mathrm{TopK}(s(h),k)} s_t(h)$, where $\mathrm{TopK}(\cdot,k)$ selects indices of the largest $k$ temporal values. The critic then combines tail-risk evidence with global average pooling (GAP) to preserve overall realism, $z=\mathrm{CVaRPool}(h,k)+\mathrm{GAP}(h)$, computed for real and fake as $z_r,z_f$ \textbf{(Rows \textcircled{\scriptsize 4}--\textcircled{\scriptsize 5}}). Finally, an Multi Layer Perceptron (MLP) maps $z_r,z_f$ to critic scores $r,f\in\mathbb{R}^{B\times 1}$ \textbf{(Row \textcircled{\scriptsize 6})}, and the forward pass returns both scores and feature maps \textbf{(Row \textcircled{\scriptsize 7}}). Hence, by design, CVaRD addresses the mean-aggregation failure mode: \textit{the feedback adversarial signal is dominated by the perceptually important HF worst-case (tail) segments rather than average frames.}

\vspace{-00.9em}
\begin{algorithm}[ht]
\caption{Chance-Constraint HF Discriminator (CCD)}
\label{alg:ccd}
\begin{algorithmic}[1]\scriptsize
\Require $y_{\mathrm{real}}, y_{\mathrm{fake}} \in \mathbb{R}^{B \times 1 \times T}$; STFT $(n_{\mathrm{fft}}, h, w)$; HF fraction $\rho$; barrier $(\kappa, \tau)$, where $(\kappa, \tau)$ are thresholds
\Ensure $[r], [f]$ and FM maps $\{F_r^{(1,2)}\}, \{F_f^{(1,2)}\}$

\State $q_r \gets \mathrm{HFRatio}(\mathrm{STFT}(\mathrm{pad}(y_{\mathrm{real}})); \rho)$ \Comment{per-frame HF/total}
\State $q_f \gets \mathrm{HFRatio}(\mathrm{STFT}(\mathrm{pad}(y_{\mathrm{fake}})); \rho)$ \Comment{same on fake}
\State $b_r \gets \log(1+\exp(\kappa(q_r-\tau)))$ \Comment{soft barrier}
\State $b_f \gets \log(1+\exp(\kappa(q_f-\tau)))$ \Comment{soft barrier}
\State $h_r \gets \mathrm{SepConvHead}(b_r)$; $h_f \gets \mathrm{SepConvHead}(b_f)$ \Comment{1D head}
\State $r \gets \mathrm{AvgPool}(h_r) + \mathrm{MaxPool}(h_r)$ \Comment{global score (avg+max)}
\State $f \gets \mathrm{AvgPool}(h_f) + \mathrm{MaxPool}(h_f)$ \Comment{global score (avg+max)}
\State $\{F_r^{(1,2)}\} \gets \{b_r, h_r\}; \quad \{F_f^{(1,2)}\} \gets \{b_f, h_f\}$ \Comment{FM maps}
\State \Return $([r], [f], \{F_r^{(1,2)}\}, \{F_f^{(1,2)}\})$ \Comment{barrier + head-out}
\end{algorithmic}
\end{algorithm}
\vspace{-1em}

\begin{table*}[ht]
\centering
\scriptsize
\setlength{\tabcolsep}{1.5pt}
\renewcommand{\arraystretch}{1}

\begin{tabular}{l l l
  ccc  ccc  ccc  ccc  ccc  ccc ccc}
\toprule
\multirow{2}{*}{Method}
& \multirow{2}{*}{Size}
& \multirow{2}{*}{Data}
  & \multicolumn{3}{c}{NISQA-MOS$\uparrow$}
  & \multicolumn{3}{c}{STOI$\uparrow$}
  & \multicolumn{3}{c}{PESQ$\uparrow$}
  & \multicolumn{3}{c}{SI-SDR$\uparrow$}
  & \multicolumn{3}{c}{SI-SNR$\uparrow$}
  & \multicolumn{3}{c}{LSD$\downarrow$} 
  & \multicolumn{3}{c}{WER \%$\downarrow$}\\
  \cmidrule(lr){4-6}\cmidrule(lr){7-9}\cmidrule(lr){10-12}
\cmidrule(lr){13-15}\cmidrule(lr){16-18}\cmidrule(lr){19-21}\cmidrule(lr){22-24}
  & &
  & 4–16 & 8–16 & 16–48
  & 4–16 & 8–16 & 16–48
  & 4–16 & 8–16 & 16–48
  & 4–16 & 8–16 & 16–48
  & 4–16 & 8–16 & 16–48
  & 4–16 & 8–16 & 16–48 
  & 4–16 & 8–16 & 16–48\\
\midrule
\multirow{2}{*} {Unprocessed}   & \multirow{2}{*} {-}     & VCTK clean& 2.79 & 3.67 & 4.43 & 0.55 & 0.61 & 0.61 & 1.15  & 1.51  & 1.41  & -11.03 & -8.07 & -6.07 & -10.53 & -7.62 & -5.63 & 3.27 & 2.27 & 2.85  & 90.0 & 6.1 & 2.1\\

&   & MLS clean & 2.29 & 3.12 & - & 0.51 & 0.57 & - & 1.05  & 1.34  & -  & -14.15 & -18.89 & - & -13.67 & -17.71 & - & 3.48 & 2.48 & - & 91.1 & 6.8 & -\\

\hline

\multirow{2}{*} {EBEN, 2023}   & \multirow{2}{*} {29.7M} & VCTK clean& 2.59 & 2.69 & 2.53     & 0.89 & 0.98 & 0.98      & \textbf{2.64}  & 3.69  & 3.71      & 11.94  & 19.94 & 20.82      & 11.94  & 19.94 & 20.83      & 1.03  & 0.78  & 0.92   & 19.1 & 12.4 & 8.5 \\
   &   & MLS clean& 2.14 & 2.28 & -     & 0.87 & 0.96 & -      & 2.47  & 3.55  & -      & 11.81  & 18.24 & -      & 11.74  & 18.38 & -      &  1.17 &  0.88 & -   & 19.4 & 12.7 & - \\

\hline

\multirow{2}{*} {AERO, 2023}  & \multirow{2}{*} {36.4M} & VCTK clean& 2.79 & 2.75 & 2.88 & 0.83 & 0.94 & 0.99  & 2.62  & 3.65  & 3.69  & \textbf{13.60}  & \textbf{20.70} & \textbf{21.56} & \textbf{13.60}  & \textbf{20.70} & \textbf{21.56} & 1.09  & 0.97  & 0.75 & 20.3  & 13.4 & 9.8\\
&   & MLS  clean& 2.38 & 2.37 & - & 0.81 & 0.92 & -  & 2.49  & 3.45  & -  &  13.47 & \textbf{19.41} & - & \textbf{13.37}  & \textbf{19.19} & - & 1.21  & 1.04  & - & 20.7  & 13.7 & -\\

\hline

\multirow{2}{*} {AP-BWE, 2024} & \multirow{2}{*} {72M}  & VCTK clean& 3.86 & 3.97 & 4.49 & 0.94 & 0.99 & 0.99  & 2.55  & \textbf{3.69}  & 3.72  & 13.42  & 18.26 & 20.86 & 13.35  & 18.07 & 20.74 & 0.96  & \textbf{0.74}  & 0.75 & \textbf{13.7} & \textbf{4.2}  & 1.7\\ 
&   & MLS clean& 3.18& 3.42 & - & 0.92&0.99  & -  & \textbf{2.51}  & \textbf{3.62}  & -  & \textbf{13.57}  & 17.39 & - & 13.17  & 16.33 & - & 1.11  &  0.84 & - & \textbf{13.9} & \textbf{4.3} & -\\  

\hline



\multirow{2}{*} {SwinDS-BWE}         & \multirow{2}{*} {18.5M} & VCTK clean & \textbf{4.24} & \textbf{4.32} & \textbf{4.5} & \textbf{0.94} & \textbf{0.99} & \textbf{0.99}  & 2.46  & {3.48}  & \textbf{4.4}  & {12.95}  & {17.04} & {19.2} & {12.98}  & {17.91} & {19.16} & \textbf{0.95}  & 0.75  & \textbf{0.74} & 14.3 & 4.9 & \textbf{1.62}\\

(our) &   & MLS clean& \textbf{3.18} & \textbf{3.54} & - & \textbf{0.92} & \textbf{0.99} & -  & 2.39  & 2.96 & -  & {11.99}  & {14.75} & - & {11.89}  & {14.65} & - & \textbf{1.09}  & \textbf{0.81}  & - & \textbf{} & \textbf{} & \textbf{-}\\

\hline

AP-BWE         & 72M & VCTK noisy & 2.74 & 3.15 & 3.71 & -- & -- & --  & \textbf{1.26}  & 1.68  & 1.45  & --  & -- & -- & \textbf{4.36}  & 5.31 & 6.11 & \textbf{1.37}  & 1.07  & 0.89 & - & - & -\\

SwinDS-BWE         & 18.5M & VCTK noisy & \textbf{2.94} & \textbf{3.98} & \textbf{4.46} & -- & -- & --  & 1.06 & \textbf{3.16} & \textbf{4.34}  & -- & -- & --  & 4.15  & \textbf{17.12} & \textbf{20.46} & 1.58 &  \textbf{0.89}  & \textbf{0.88} & - & - & -\\




\bottomrule
\end{tabular}
\vspace{-1.11em}
\caption{Comparison for clean English-VCTK and MLS French dataset (MLS does not have 48 kHz speech).}
\label{tab:Comparative_analysis}
\vspace{-02.0em}
\end{table*}

\vspace{-01.20em}
\subsubsection{\textbf{Chance-Constraint HF Discriminator (CCD): For barrier trace to suppress over-boost (see Algorithm~\ref{alg:ccd})}}
\label{subsec:disc_cc}
\vspace{-0.41em}

Discriminators can be fooled by spurious HF overshoot, producing perceptually ``bright'' yet unrealistic artifacts. The CCD mitigates this by acting as a \emph{regularizer} that converts a physically interpretable HF-risk criterion \cite{charnes1959chance} (a chance-style constraint on per-frame HF energy ratio) into a differentiable barrier utility used for adversarial supervision and feature matching. It computes an HF-ratio trace $q[\ell]=\mathrm{HFRatio}(\mathrm{STFT}(\cdot);\rho)$ for real and fake (\textbf{Rows \textcircled{\scriptsize 1}--\textcircled{\scriptsize 2}}), where $\rho\in(0,1)$ specifies the HF portion and $\tau$ (=0.4) is a predefined HF-ratio threshold, and STFT =  Short-Time Fourier Transform. A smooth barrier $b[\ell]=\log \left( 1 + \exp \left( \kappa \big( q[\ell] - \tau \big) \right) \right)$ is then applied (\textbf{Rows \textcircled{\scriptsize 3}--\textcircled{\scriptsize 4}}), which is exponentially small for $q[\ell]<\tau$ and becomes approximately linear in $(q[\ell]-\tau)$ for $q[\ell]>\tau$ where sharpness is controlled by $\kappa>0$. The barrier sequence is processed by a SepConv-1D head to produce $h_r,h_f$ (\textbf{Rows \textcircled{\scriptsize 5}}), and scores are formed via global max pooling (MaxPool(.)) and global average pooling (AvgPool(.)) in \textbf{Rows \textcircled{\scriptsize 6}--\textcircled{\scriptsize 7}} for $r$ and $f$, 
so persistent overshoot increases the average while rare spikes are captured by the maximum. For feature matching, CCD returns two intermediate maps $F^{(1)}=b$ (barrier trace) and $F^{(2)}=h$ (head activation) (\textbf{Row \textcircled{\scriptsize 8}}), alongside final scores $([r],[f])$ (\textbf{Row \textcircled{\scriptsize 9}}).

\vspace{-0.71em}
\subsubsection{\textbf{Multi-Criteria Utility Discriminator (MCUD): For criteria-to-utility aggregation for HF realism (see Algorithm~\ref{alg:mcud})}}
\label{subsubsec:disc_mcu}
\vspace{-0.31em}

HF realism is a multi-objective trade-off (e.g., brightness vs.\ noisiness, tonality vs.\ spectral spread), yet traditional discriminators overlook this. MCUD explicitly models this trade-off structure by extracting interpretable STFT criteria and aggregating them with a learnable \emph{convex} utility \cite{keeney1993decisions}. MCUD first computes magnitude spectra $A_r,A_f=|\mathrm{STFT}(\cdot)|+\varepsilon$ (\textbf{Row \textcircled{\scriptsize 1}}), then extracts per-frame criterion, $C=\{\mathrm{SF},\mathrm{SC},\mathrm{SR}_{0.85}\}$ (\textbf{Row \textcircled{\scriptsize 2}}), where SF=Spectral Flatness, SC=Spectral Centroid, and SR=Spectral Roll with a roll-off $p{=}0.85$. The Instance-Normalization (IN) normalizes each criterion ($C$) over time to \emph{focus on temporal patterns}  (\textbf{Row \textcircled{\scriptsize 3}}). Then, MCUD learns simplex weights $w=\mathrm{softmax}(\theta)$ (\textbf{Row \textcircled{\scriptsize 4}}) with $w_i\ge 0$ and $\sum_i w_i=1$, and forms a per-frame utility trace by convex aggregation, $u(\tau)=w^\top C_n(:,\tau)$, producing $u_r(\tau),u_f(\tau)$ for real and fake (\textbf{Row \textcircled{\scriptsize 5}}). A SepConv + GAP scoring head maps the utility traces to scalar hinge logits $r,f\in\mathbb{R}^{B\times 1}$ (\textbf{Row \textcircled{\scriptsize 6}}). For feature matching, MCUD returns two intermediate maps: $F^{(1)}=C_n$ (normalized criteria stack) and $F^{(2)}=u$ (utility trace), i.e., $\{F_r^{(1,2)}\}=\{C_{n,r},u_r\}$ and $\{F_f^{(1,2)}\}=\{C_{n,f},u_f\}$ (\textbf{Row \textcircled{\scriptsize 7}}), alongside the final scores $([r],[f])$ (\textbf{Row \textcircled{\scriptsize 8}}). MCUD provides an interpretable critic that learns how to balance multiple HF criteria to separate real vs.\ fake with minimal overhead.

\vspace{-0.70em}
\begin{algorithm}[ht!]
\caption{Multi-Criteria Utility Discriminator (MCUD)}
\label{alg:mcud}
\begin{algorithmic}[1]\scriptsize
\Require $y_{\mathrm{real}}, y_{\mathrm{fake}} \in \mathbb{R}^{B \times 1 \times T}$; STFT $(n_{\mathrm{fft}}, h, w)$; sample rate $f_s$
\Ensure $[r], [f]$ and FM maps $\{F_r^{(1,2)}\}, \{F_f^{(1,2)}\}$

\State $A_r \gets |\mathrm{STFT}(\mathrm{pad}(y_{\mathrm{real}}))| + \varepsilon; \quad A_f \gets |\mathrm{STFT}(\mathrm{pad}(y_{\mathrm{fake}}))| + \varepsilon$ 
\State $C_r \gets \mathrm{Crit}(A_r) = \{\mathrm{SF}, \mathrm{SC}, \mathrm{SR}_{0.85}\}; \quad C_f \gets \mathrm{Crit}(A_f)$ 
\State $C_{n,r} \gets \mathrm{IN}(C_r); \quad C_{n,f} \gets \mathrm{IN}(C_f)$ \Comment{instance-norm over time}
\State $w \gets \mathrm{softmax}(\theta)$ \Comment{simplex weights (learned)}
\State $u_r(\tau) \gets w^\top C_{n,r}(:, \tau); \quad u_f(\tau) \gets w^\top C_{n,f}(:, \tau)$ \Comment{Utility trace}
\State $r \gets \mathrm{Head}(u_r); \quad f \gets \mathrm{Head}(u_f)$ \Comment{SepConv $\to$ GAP $\to$ Linear}
\State $\{F_r^{(1,2)}\} \gets \{C_{n,r}, u_r\}; \quad \{F_f^{(1,2)}\} \gets \{C_{n,f}, u_f\}$ \Comment{FM maps}
\State \Return $([r], [f], \{F_r^{(1,2)}\}, \{F_f^{(1,2)}\})$ \Comment{criteria + utility}
\end{algorithmic}
\end{algorithm}

\vspace{-02.0em} 
\subsection{Loss Functions}
\vspace{-0.52em}

We use the same six losses from \cite{lu2024towards} for generator loss. The total discriminator loss $\mathcal{L}_D$ is shown in Eqn. \ref{eqn:discriminitorloss}, where $\mathcal{L}_D^{\text{CVaRD}}$, $\mathcal{L}_D^{\text{CCD}}$, $\mathcal{L}_D^{\text{MCUD}}$, $\mathcal{L}_D^{\text{MRAD}}$, and $\mathcal{L}_D^{\text{MRPD}}$ are CVaRD, CCD, MCUD, Multi-Resolution Amplitude (MRAD) \cite{lu2024towards}, and Multi-Resolution Phase Discriminator (MRPD) \cite{lu2024towards} losses, respectively. 

\vspace{-1.1em}
{
\scriptsize
\begin{equation}
\mathcal{L}_D = \sum_r \mathcal{L}_D^{\text{CVaRD}}  + 
\sum_r \mathcal{L}_D^{\text{CCD}}  + \sum_r \mathcal{L}_D^{\text{MCUD}} 
\sum_r \mathcal{L}_D^{\text{MRAD}}   + 
\sum_r \mathcal{L}_D^{\text{MRPD}}   
\label{eqn:discriminitorloss}
\end{equation}
}

\vspace{-01.70em}
\section{Analysis, Results, and Ablation Study}
\label{sec:results}
\vspace{-0.3em}


 \textbf{Corpus and preprocessing:} We use the VCTK (v0.92)~\cite{sarfjoo2018} (110 speakers) and MLS French corpus \cite{Pratap2020MLSAL} (114 speakers). 
 We load 16/48 kHz files, convert to mono channel, remove silence parts, downsample to simulate band-limited audio, sinc interpolate, and align segments.

\textbf{Optimization and hyperparameters:} We train with AdamW (learning rate $2\times 10^{-4}$, $\beta_1=0.8$, $\beta_2=0.99$), exponential LR decay $\gamma=0.999$, gradient clipping 10.0, batch size 16, and seed 1234. Swin-1D uses a window size $W=8$ and alternating shift $W/2$. Decision-science hyperparameters follow the implementation: CVaR tail fraction $\alpha=0.2$; chance-constraint HF ratio fraction $\rho=0.25$, barrier $\kappa=8$, threshold $\tau_0=0.40$; multi-criteria uses SF/SC/SR with learnable convex weights. Models are trained for 50 epochs on a NVIDIA RTX~4090 GPU and Intel Xeon Silver~4310 CPUs.


\textbf{Metrics:} We use Log-Spectral Distance (LSD) for spectral distance, Short-Time Objective Intelligibility (STOI) for intelligibility, Perceptual Evaluation of Speech Quality (PESQ) for quality, Scale-Invariant Signal-to-Distortion Ratio (SI-SDR) and Scale-Invariant Signal-to-Noise Ratio (SI-SNR) for distortion, Non-Intrusive Speech Quality Assessment - Mean Opinion Score (NISQA-MOS) \cite{mittag2021nisqa}  for opinion score, and Word Error Rate (WER) for  ASR task.

\vspace{-1.32em}
\subsection{Comparative Analysis with Baselines with \& without Noise}
\label{subsec:comparative-baselines}
\vspace{-0.5em}

Across the three frequency ranges (4–16, 8–16, 16–48\,kHz), our proposed SwinDS-BWE consistently delivers the \textbf{best speech quality (NISQA-MOS)} while maintaining \emph{almost similar intelligibility (STOI)} and LSD (see Table \ref{tab:Comparative_analysis}). Specifically, SwinDS-BWE provides the highest NISQA-MOS in all three frequency ranges, outperforming EBEN, AERO, and AP-BWE. STOI is also higher for SwinDS-BWE than EBEN/AERO and on par with AP-BWE. In terms of PESQ, SwinDS-BWE performs far better than all the other baselines for 16-48 kHz range, but has slightly reduced PESQ for the 4-16 and 8-16 kHz ranges. In terms of LSD, SwinDS-BWE outperforms EBEN \& AERO but is similar to AP-BWE. However, SwinDS-BWE has reduced {\color{black}SI-SDR and SI-SNR} compared to AP-BWE, but has mixed superiority across different frequency ranges over EBEN \& AERO. However, though AP-BWE, EBEN \& AERO exhibit slightly higher SI-SDR/SI-SNR at highband fills, but \textbf{they lag substantially in NISQA-MOS, STOI, and WER compared to SwinDS-BWE}, suggesting that SwinDS-BWE's signal fidelity is much better translated to human-perceived quality compared to other recent baselines. 

We also show that SwinDS-BWE outperforms AP-BWE in NISQA-MOS, SI-SNR, and LSD for the noisy VCTK dataset. We use 8 noise sources from AURORA dataset \cite{pearce00_icslp} and five SNR levels ($-10$, $-5$, $0$, $5$, $10)$ dB. 
The selected noises cover both stationary and non-stationary conditions, including airport, babble, car, exhibition, station, street, speech-shaped noise, and white Gaussian noise.


\textbf{Parameter efficiency:} 
We compare SwinDS-BWE with the best-performing AP-BWE. SwinDS-BWE achieves the highest NISQA-MOS and STOI, and similar LSD with \textit{only 18.45M parameters, which is 74\% lower than AP-BWE’s 72M parameter count.} Given the higher NISQA-MOS at comparable STOI, PESQ, and LSD, this indicates a favorable accuracy–capacity balance. The parameter reduction happens with a better/similar performance gain because of our newly designed decision-science discriminators -- CVaRD, CCD, and MCUD -- combining with the Swin-1D SW-MSA stack. 


\vspace{-0.81em}
\subsection{Ablation Study}
\label{subsec:ablation}
\vspace{-0.41em}

To verify that the proposed \emph{decision-science} discriminators contribute more than just a generic adversarial penalty, we design an ablation study in Table \ref{table:swinDS_discriminator_ablation}. 
We hypothesize that BWE errors are \emph{heteroscedastic in time}: most frames are reconstructed cleanly, while perceptually dominant failures cluster around short HF-critical events, resulting in a reduction in perceptual subjective quality. 

\vspace{-0.8em}
\begin{table}[ht!]
  \centering
  \scriptsize
  \setlength{\tabcolsep}{1.6pt}
\begin{tabular}{l c c c c c c c c c c}
\toprule
SL & MRAD & MRPD & CVaRD & CCD & MCUD & LSD & STOI & PESQ & SNR & N-MOS \\[-0.2ex]
\hline
\multicolumn{11}{c}{}\\[-6.5pt]
\multicolumn{11}{c}{\textbf{Addition of Single Decision-Science Discriminators}}\\[-0.4pt] \hline
1 & \xmark & \xmark & \cmark & \xmark & \xmark &  1.06 & 0.94 & 2.32 & 12.89 & 3.36  \\
2 & \xmark & \xmark & \xmark & \cmark & \xmark &  1.25 & 0.95 & 2.61 & 14.35 & 2.99  \\
3 & \xmark & \xmark & \xmark & \xmark & \cmark &  1.18 & 0.95 & 2.63 & 14.2  & 3.01  \\\hline
\multicolumn{11}{c}{\textbf{Decision Science Discriminators with MRAD and MRPD Pair}}\\\hline
4 & \xmark & \xmark & \cmark & \cmark & \cmark &  1.06 & 0.94 & 2.27 & 11.8  & 3.23  \\
5 & \cmark & \xmark & \cmark & \cmark & \cmark &  1.04 & 0.94 & 2.27 & 12.94 & 4.04  \\
6 & \xmark & \cmark & \cmark & \cmark & \cmark &  1.02 & 0.94 & 2.29 & 12.77 & 4.16   \\\hline
\multicolumn{11}{c}{\textbf{Comparision of Swin-1D vs ConvNeXt \cite{lu2024towards} in the Generator}}\\[-0.4pt] \hline
7 & \cmark & \cmark & \cmark & \cmark & \cmark &  1.0 & 0.94  & 2.25 & 12.13 & 4.13  \\
\rowcolor{green!25}
8 & \cmark & \cmark & \cmark & \cmark & \cmark &  1.03 & 0.93 & 2.2  & 12.57 & 4.24  \\
\bottomrule
\end{tabular}
\vspace{-0.9em}
\caption{Ablation of decision-science discriminators for 4$\rightarrow$16~kHz. Here, N-MOS = NISQA-MOS and SNR = SI-SNR.}
\label{table:swinDS_discriminator_ablation}
\vspace{-01.0em}
\end{table}

\textbf{(i) Single decision-science discriminators:} We add CVaRD, CCD, and MCUD one at a time without MRAD and MRPD in \textbf{Rows \textcircled{\scriptsize 1}-\textcircled{\scriptsize 3}}. CVaRD in \textbf{Row \textcircled{\scriptsize 1}} improves HF realism by concentrating updates on the \emph{worst} short events (tail-risk frames), yielding stronger perceptual gains than a pure “average-frame” adversarial signal \emph{(highest NISQA-MOS = 3.36, versus 2.99 for CCD and 3.01 for MCUD; i.e., $\Delta$NISQA-MOS = +0.37/+0.35, with LSD = 1.06 vs 1.25/1.18)}.  CCD in \textbf{Row \textcircled{\scriptsize 2}} acts as an HF-energy ratio regulator. It most directly suppresses persistent HF over-activation (hiss-like boosting) by aligning HF-energy dynamics to a feasible region, typically improving energy-based scores but not necessarily maximizing perceived HF naturalness when used alone \emph{(SI-SNR = 14.35~dB and PESQ = 2.61 are the best among Rows 1--3, while NISQA-MOS is the lowest at 2.99)}. MCUD in \textbf{Row \textcircled{\scriptsize 3}}  behaves as a texture selector. By learning a convex utility over flatness/centroid/rolloff cues, it stabilizes HF ``grain'' and reduces over-smoothed artifacts that are not captured by any single spectral descriptor \emph{(PESQ = 2.63 is best among Rows 1--3, with NISQA-MOS = 3.01)}. 

\textbf{(ii) Decision-science discriminators with MRAD and MRPD pair:} We evaluate our model with only three decision-science discriminators in \textbf{Row \textcircled{\scriptsize 4}} and separately add MRAD in \textbf{Row \textcircled{\scriptsize 5}} and MRPD in \textbf{Row \textcircled{\scriptsize 6}}. MRAD or MRPD produces the clearest NISQA-MOS improvement \emph{(3.23 $\rightarrow$ 4.04 / 4.16, i.e., $\Delta$NISQA-MOS = +0.81 / +0.93)}, supporting the interpretation of the \emph{global structure first, local HF polishing complimentary}. Concurrently, LSD improves from 1.06 $\rightarrow$ 1.04 / 1.02, and SI-SNR improves from 11.8 $\rightarrow$ 12.94 / 12.77~dB. In practice, MRAD/MRPD stabilizes reconstruction, after which CVaRD/CCD/MCUD can allocate capacity to the sparse HF-critical failures rather than compensating for global mismatches. Hence, we can claim that three decision science discriminators act as \emph{expert in-painter} for the sparse HF contents. 

\textbf{(iii) Comparison of Swin-1D vs ConvNeXt \cite{lu2024towards} in the generator:} With the choice of discriminator set fixed, the generator choice mainly changes perceptual trade-offs. We include \textbf{Row \textcircled{\scriptsize 7}} as a reference, which replaces Swin-1D with the ConvNeXt block from AP-BWE's generator under the full discriminator set, and \textbf{Row \textcircled{\scriptsize 8}} as the corresponding Swin-1D generator counterpart. The NISQA-MOS improves from 4.13 $\rightarrow$ 4.24, while PESQ changes from 2.25 $\rightarrow$ 2.20, and LSD changes from 1.00 $\rightarrow$ 1.03, SNR improves from 12.13 $\rightarrow$ 12.57~dB. \textit{The Swin generator improves perceived quality (higher NISQA-MOS) while using the same discriminator feedback with 1.9x fewer parameters, indicating that the proposed Swin-1D Lattice generator is architecturally preferred for producing more natural-sounding audios from the same discriminator-side feedback.}

\vspace{-1.2em} 
\subsection{Results Across Different Bands} 
\vspace{-0.4em}


Table \ref{table:freqrange} shows a consistent trend across all objective metrics: \emph{as the reconstruction band becomes less demanding (i.e., the gap between the narrow and target bands shrinks), performance improves}. The broadest setting (2-48 kHz) is the most difficult, yielding the weakest scores. As the effective gap is reduced, distortion drops and intelligibility/quality rises. 

\vspace{-01.92em}
\begin{table}[ht]
\caption{Performance over different frequency ranges.}
\vspace{-0.1em}
\centering
\scriptsize
\setlength{\tabcolsep}{3.5pt}
\renewcommand{\arraystretch}{0.85}

\begin{tabular}{lcccccc}
\toprule
Freq. range &  LSD & STOI  & PESQ   & SI-SDR & SI-SNR & NISQA-MOS\\ 
\midrule
2-16 kHz   & 1.11 & 0.87  & 1.61 & 7.63  & 7.62   & 4.13\\  
2-48 kHz   & 1.13 & 0.83  & 1.19 & 7.6   & 7.59   & 4.01\\  
4-16 kHz   & 0.99 & 0.94  & 2.34 & 12.76 & 12.68  & 4.11\\  
8-16 kHz   & 0.77 & 0.998 & 3.69 & 17.59 & 17.43  & 4.29\\  
8-48 kHz  & 0.93 & 0.996 & 2.42 & 15.53 & 15.42  & 4.52\\  
12-48 kHz & 0.85 & 0.998 & 3.21 & 18.01 & 17.91  & 4.51\\  
16-48 kHz  & 0.79 & 0.998 & 3.64 & 19.5  & 19.44  & 4.50\\  
24-48 kHz  & 0.65 & 0.999 & 4.16 & 22.67 & 22.66  & 4.52 \\ 
\bottomrule
\end{tabular}
\vspace{-1.999em}
\label{table:freqrange}
\end{table}

\vspace{-00.999em}
\subsection{Computational Complexity}
\label{subsec:computational-complexity}
\vspace{-0.4em}

Table~\ref{tab:computation} summarizes the computational cost in terms of generator/discriminator parameters (GP/DP),  Multiply Accumulate Operations (MACs) and Floating Point Operations per second (FLOPs) in millions (M), and real-time factor (RTF). 
Relative to AP-BWE, SwinDS-BWE achieves a 42.6\% reduction in GP and a 96.8\% reduction in DP, and a 63\% reduction in MACS and FLOPs. However, its RTF is slightly higher compared to AP-BWE (though still real-time) because of the windowed/shifted attention in terms of window partitioning/reversal, masking, etc.

\vspace{-2.299em}
\begin{table}[ht]
\caption{Computational complexity of SwinDS-BWE.}
\centering
\scriptsize
\setlength{\tabcolsep}{1.3pt}
\renewcommand{\arraystretch}{0.85}
\begin{tabular}{lcccccc}
\toprule
Model & Fq. Range & GP (M) & DP (M) & MACs (M) & FLOPs (M) & RTF (GPU) \\
\midrule
\rowcolor{red!30}
AP-BWE & 4-16 kHz  & 29.76 & 42.3 & 14236.65 & 28473.31   & 0.0023x \\
\rowcolor{red!30}
AP-BWE & 16-48 kHz & 29.76 & 42.3 & 14236.65 & 28473.31   & 0.0025x \\
\rowcolor{blue!30}
SwinDS-BWE & 4-16 kHz  & 17.09 & 1.36 & 5277.35  & 10554.69   & 0.0032x \\
\rowcolor{blue!30}
SwinDS-BWE & 16-48 kHz & 17.09 & 1.36 & 5277.35  & 10554.69   & 0.0034x \\
\bottomrule
\end{tabular}
\vspace{-0.74em}
\label{tab:computation}
\vspace{-1.4em}
\end{table}

\vspace{-01.29em}
\subsection{Subjective Analysis}
\label{subsec:subjective-test}
\vspace{-0.5em}

Subjective comparison of SwinDS-BWE against AP-BWE and unprocessed audio is conducted by a selected panel of 10 persons. We use 5-point (1=bad to 5=excellent) Mean Opinion Score (MOS) ratings for the subjective evaluation. 
 In the bottom-right of Fig. \ref{fig:overall_architecture}, we show that SwinDS-BWE significantly outperforms AP-BWE for female speakers and overall.

\vspace{-01.51em}
\section{Conclusion and Limitation}
\vspace{-0.84em}

We design SwinDS-BWE, a bandwidth extension framework that targets two deployment-relevant bottlenecks: efficient generation and \textbf{HF-focused supervision}. The proposed \textbf{Swin-1D shifted-window AP lattice} preserves amplitude--phase coupling while replacing heavier refiners with locality-aware attention, reducing the generator to \textbf{17.09 M} parameters and cutting compute from \textbf{14.24 G} to \textbf{5.28 G} MACs while maintaining strong performance across extension bands.

To better match perceptually dominant and sparse HF errors, we introduce three lightweight \textbf{decision-science} critics: \textbf{CVaRD} (tail pooling for worst-case segments), \textbf{CCD} (chance-constraint barrier to suppress HF over-boost), and \textbf{MCUD} (convex utility over interpretable spectral criteria) with a small overhead. 
\textit{Overall, SwinDS-BWE shifts the BWE training paradigm from mean-frame matching toward tail-event realism with 74\% reduction in the parameter count.} 
However, the current design does not use data-driven calibration of hyperparameters in CCD, MCUD, and CVaRD. 

\vspace{-0.900em}
\section{Compliance with Ethical Standards and Acknowledgment}
\vspace{-0.3em}

We acknowledge the use of \textit{Elicit} and \textit{ChatGPT} for code debugging and proofreading. This work is supported by the Office of Naval Research (ONR) with an award number 000142612163.




\bibliographystyle{IEEEbib}
\bibliography{mybib}

@inproceedings{pearce00_icslp,
  author    = {David Pearce and Hans-G{\"u}nter Hirsch},
  title     = {The Aurora Experimental Framework for the Performance Evaluation of Speech Recognition Systems under Noisy Conditions},
  booktitle = {Proceedings of the 6th International Conference on Spoken Language Processing (ICSLP 2000)},
  year      = {2000},
  pages     = {29--32},
  doi       = {10.21437/ICSLP.2000-743},
  publisher = {ISCA Archive},
}

@article{Pratap2020MLSAL,
  title        = {MLS: {A} Large-Scale Multilingual Dataset for Speech Research},
  author       = {Vineel Pratap and Qiantong Xu and Anuroop Sriram and Gabriel Synnaeve and Ronan Collobert},
  journal      = {arXiv preprint arXiv:2012.03411},
  year         = {2020},
  url          = {https://arxiv.org/abs/2012.03411},
}

@inproceedings{durugkar2017generative,
  title={Generative Multi-Adversarial Networks},
  author={Durugkar, Ishan P. and Gemp, Ian and Mahadevan, Sridhar},
  booktitle={ICLR Workshop},
  year={2017}
}

@inproceedings{tamiti-etal-2026-cis,
    title = "{CIS}-{BWE}: Chaos-Informed Speech Bandwidth Extension",
    author = "Tamiti, Tarikul Islam  and
      Das, Tonmoy  and
      Mamun, Nursadul  and
      Barua, Anomadarshi",
    booktitle = "Proc. {ACL} (Volume 1: Long Papers)",
    year = "2026",
    url = "https://aclanthology.org/2026.acl-long.1925/",
    doi = "10.18653/v1/2026.acl-long.1925",
    pages = "41509--41529",
    ISBN = "979-8-89176-390-6"
}

@INPROCEEDINGS{11464384,
  author={Tamiti, Tarikul Islam and Barua, Anomadarshi},
  booktitle={Proc. {ICASSP}},
  title={NLDSI-BWE: Non Linear Dynamical Systems-Inspired Multi Resolution Discriminators for Speech Bandwidth Extension}, 
  year={2026},
  volume={},
  number={},
  pages={18907-18911},
  doi={10.1109/ICASSP55912.2026.11464384}}

@article{lu2024towards,
  title={Towards high-quality and efficient speech bandwidth extension with parallel amplitude and phase prediction},
  author={Lu, Ye-Xin and Ai, Yang and Du, Hui-Peng and Ling, Zhen-Hua},
  journal={IEEE/ACM Transactions on Audio, Speech, and Language Processing},
  year={2024},
  publisher={IEEE}
}

@inproceedings{liu2021swin,
  title={Swin transformer: Hierarchical vision transformer using shifted windows},
  author={Liu, Ze and Lin, Yutong and Cao, Yue and Hu, Han and Wei, Yixuan and Zhang, Zheng and Lin, Stephen and Guo, Baining},
  booktitle={Proceedings of the IEEE/CVF international conference on computer vision},
  pages={10012--10022},
  year={2021}
}

@inproceedings{kim2023hifi++,
  title={{HiFi++}: Towards Perceptually Enhanced and Computationally Efficient Neural Vocoders},
  author={Kim, Woosung and Kang, Hyeongmin and Kim, Younggun and Jung, Kyunggu and Lee, Joon Son and Lee, Sang-Hoon},
  booktitle={Proc. Interspeech},
  year={2023},
  pages={4374--4378}
}

@inproceedings{liu2022neural,
  title={Neural vocoder is all you need for speech super-resolution},
  author={Liu, Haohe and Choi, Woosung and Liu, Xubo and Kong, Qiuqiang and Tian, Qiao and Wang, DeLiang},
  year      = {2022},
  booktitle = {Interspeech},
}

@article{sui2024tramba,
  title={Tramba: A hybrid transformer and mamba architecture for practical audio and bone conduction speech super resolution and enhancement on mobile and wearable platforms},
  author={Sui, Yueyuan and Zhao, Minghui and Xia, Junxi and Jiang, Xiaofan and Xia, Stephen},
  journal={Proceedings of the ACM on Interactive, Mobile, Wearable and Ubiquitous Technologies},
  volume={8},
  number={4},
  pages={1--29},
  year={2024},
  publisher={ACM New York, NY, USA}
}

@article{tagliasacchi2020seanet,
  title={SEANet: A multi-modal speech enhancement network},
  author={Tagliasacchi, Marco and Li, Yunpeng and Misiunas, Karolis and Roblek, Dominik},
  journal={arXiv preprint arXiv:2009.02095},
  year={2020}
}

@inproceedings{luo2020latticenet,
  title={Latticenet: Towards lightweight image super-resolution with lattice block},
  author={Luo, Xiaotong and Xie, Yuan and Zhang, Yulun and Qu, Yanyun and Li, Cuihua and Fu, Yun},
  booktitle={European conference on computer vision},
  pages={272--289},
  year={2020},
  organization={Springer}
}

@inproceedings{liu2022convnet,
  title={A convnet for the 2020s},
  author={Liu, Zhuang and Mao, Hanzi and Wu, Chao-Yuan and Feichtenhofer, Christoph and Darrell, Trevor and Xie, Saining},
  booktitle={Proceedings of the IEEE/CVF conference on computer vision and pattern recognition},
  pages={11976--11986},
  year={2022}
}

@inproceedings{uryasev2000conditional,
  title={Conditional value-at-risk: Optimization algorithms and applications},
  author={Uryasev, Stanislav},
  booktitle={proceedings of the IEEE/IAFE/INFORMS 2000 conference on computational intelligence for financial engineering (CIFEr)(Cat. No. 00TH8520)},
  pages={49--57},
  year={2000},
  organization={IEEE}
}

@article{charnes1959chance,
  title={Chance-constrained programming},
  author={Charnes, Abraham and Cooper, William W},
  journal={Management science},
  volume={6},
  number={1},
  pages={73--79},
  year={1959},
  publisher={INFORMS}
}

@book{keeney1993decisions,
  title={Decisions with multiple objectives: preferences and value trade-offs},
  author={Keeney, Ralph L and Raiffa, Howard},
  year={1993},
  publisher={Cambridge university press}
}

@article{mittag2021nisqa,
  title={NISQA: A deep CNN-self-attention model for multidimensional speech quality prediction with crowdsourced datasets},
  author={Mittag, Gabriel and Naderi, Babak and Chehadi, Assmaa and M{\"o}ller, Sebastian},
  journal={arXiv preprint arXiv:2104.09494},
  year={2021}
}

@inproceedings{yin2020phasen,
  title={Phasen: A phase-and-harmonics-aware speech enhancement network},
  author={Yin, Dacheng and Luo, Chong and Xiong, Zhiwei and Zeng, Wenjun},
  booktitle={Proceedings of the AAAI conference on artificial intelligence},
  volume={34},
  pages={9458--9465},
  year={2020}
}

@inproceedings{sfeclis2023investigating,
  title={Investigating imaginary mask estimation in complex masking for speech enhancement},
  author={Sfeclis, Georgiana-Elena and Milner, Ben and Websdale, Danny},
  booktitle={2023 31st European Signal Processing Conference (EUSIPCO)},
  pages={131--135},
  year={2023},
  organization={IEEE}
}

@misc{sarfjoo2018,
  author       = {Sarfjoo, Seyyed Saeed and Yamagishi, Junichi},
  title        = {Device Recorded VCTK (Small subset version) [sound]},
  year         = {2018},
  publisher    = {University of Edinburgh, School of Informatics, The Centre for Speech Technology Research (CSTR)},
  doi          = {10.7488/ds/2316},
  url          = {https://doi.org/10.7488/ds/2316},
  note         = {Dataset}
}

@article{lu2024multi,
  title={Multi-Stage Speech Bandwidth Extension with Flexible Sampling Rate Control},
  author={Lu, Ye-Xin and Ai, Yang and Sheng, Zheng-Yan and Ling, Zhen-Hua},
  journal={arXiv preprint arXiv:2406.02250},
  year={2024}
}

@inproceedings{su2021bandwidth,
  title={Bandwidth extension is all you need},
  author={Su, Jiaqi and Wang, Yunyun and Finkelstein, Adam and Jin, Zeyu},
  booktitle={ICASSP 2021-2021 IEEE International Conference on Acoustics, Speech and Signal Processing (ICASSP)},
  pages={696--700},
  year={2021},
  organization={IEEE}
}

@inproceedings{hauret2023eben,
  title={EBEN: Extreme Bandwidth Extension Network applied to speech signals captured with noise-resilient body-conduction microphones},
  author={Hauret, Julien and Joubaud, Thomas and Zimpfer, V{\'e}ronique and Bavu, {\'E}ric},
  booktitle={ICASSP 2023 - 2023 IEEE International Conference on Acoustics, Speech and Signal Processing (ICASSP)},
  pages={1--5},
  year={2023},
  organization={IEEE},
  doi={10.1109/ICASSP49357.2023.10096301}
}

@inproceedings{mandel2023aero,
  title={Aero: Audio super resolution in the spectral domain},
  author={Mandel, Moshe and Tal, Or and Adi, Yossi},
  booktitle={ICASSP 2023-2023 IEEE International Conference on Acoustics, Speech and Signal Processing (ICASSP)},
  pages={1--5},
  year={2023},
  organization={IEEE}
}

@article{tamiti2025high,
  title={A high-fidelity speech super resolution network using a complex global attention module with spectro-temporal loss},
  author={Tamiti, Tarikul Islam and Joshi, Biraj and Hasan, Rida and Hasan, Rashedul and Athay, Taieba and Mamun, Nursad and Barua, Anomadarshi},
  journal={arXiv preprint arXiv:2507.00229},
  year={2025}
}

@article{miyato2018spectral,
  title={Spectral normalization for generative adversarial networks},
  author={Miyato, Takeru and Kataoka, Toshiki and Koyama, Masanori and Yoshida, Yuichi},
  journal={arXiv preprint arXiv:1802.05957},
  year={2018}
}

@article{howard2017mobilenets,
  title={Mobilenets: Efficient convolutional neural networks for mobile vision applications},
  author={Howard, Andrew G and Zhu, Menglong and Chen, Bo and Kalenichenko, Dmitry and Wang, Weijun and Weyand, Tobias and Andreetto, Marco and Adam, Hartwig},
  journal={arXiv preprint arXiv:1704.04861},
  year={2017}
}

\end{document}